\documentclass[epj]{svjour}
\usepackage{graphics}
\usepackage{graphicx}
\begin{document}
\title{First measurement of the $\rho$ spectral function in nuclear collisions}
\author{S. Damjanovic\inst{1} for the NA60 Collaboration 
}                     
%
\institute{CERN, Geneva, Switzerland}
%
%
\date{Received: date / Revised version: date}
%

\abstract{ The NA60 experiment at the CERN SPS has studied low-mass
muon pairs in 158 AGeV In-In collisions. A strong excess of pairs is
observed above the yield expected from neutral meson decays. The
unprecedented sample size close to 400K events and the good mass
resolution of about 2\% made it possible to isolate the excess by
subtraction of the decay sources. The shape of the resulting mass
spectrum shows some non-trivial centrality dependence, but is largely
consistent with a dominant contribution from
$\pi^{+}\pi^{-}\rightarrow\rho\rightarrow\mu^{+}\mu^{-}$
annihilation. The associated $\rho$ spectral function exhibits
considerable broadening, but essentially no shift in mass. The
$p_{T}$-differential mass spectra show the excess to be much stronger
at low $p_{T}$ than at high $p_{T}$. The results are compared to
theoretical model predictions; they tend to rule out models linking
hadron masses directly to the chiral condensate.
} 
\authorrunning{S. Damjanovic for the NA60 Collaboration}
\titlerunning{First measurement of the $\rho$ spectral function in nuclear collisions}
\maketitle
\section{Introduction}
\label{intro}
Thermal dilepton production in the mass region $<$1 GeV/c$^{2}$ is
largely mediated by the light vector mesons $\rho$, $\omega$ and
$\phi$. Among these, the $\rho$(770 MeV/c$^{2}$) is the most
important, due to its strong coupling to the $\pi\pi$ channel and its
short lifetime of only 1.3 fm/c, much shorter than the lifetime of the
fireball. These properties have given it a key role as {\it the} test
particle for ``in-medium modifications'' of hadron properties close to
the QCD phase boundary. Changes both in width and in mass were
originally suggested as precursor signatures of the chiral
transition~\cite{Pisarski:mq}. There seems to be some consensus now
that the {\it width} of the $\rho$ should increase towards the
transition region, based on a number of quite different theoretical
approaches~\cite{Pisarski:mq,Dominguez:1992dw,Pisarski:1995xu,Rapp:1995zy,Rapp:1999ej}. On
the other hand, no consensus exists on how the {\it mass} of the
$\rho$ should change in approaching the transition: predictions exist
for a
decrease~\cite{Pisarski:mq,Brown:kk,Brown:2001nh,Hatsuda:1991ez}, a
constant behavior~\cite{Rapp:1995zy,Rapp:1999ej}, and even an
increase~\cite{Pisarski:1995xu}.

Experimentally, low-mass electron pair production was previously
investigated at the CERN SPS by the CERES /NA45 experiment for
p-Be/Au, S-Au and Pb-Au
collisions~\cite{Agakichiev:mv,Agakichiev:1995xb,Agakichiev:1997au}. The
common feature of all results from nuclear collisions was an excess of
the observed dilepton yield above the expected electromagnetic decays
of neutral mesons, by a factor of 2-3, for masses above
0.2~GeV/c$^{2}$. The surplus yield has generally been attributed to
direct thermal radiation from the fireball, dominated by pion
annihilation $\pi^{+}\pi^{-}\rightarrow\rho\rightarrow l^{+}l^{-}$
with an intermediate $\rho$ which is strongly modified by the
medium. Statistical accuracy and mass resolution of the data were,
however, not sufficient to reach the sensitivity required to assess in
detail the {\it character} of the in-medium changes.  The new
experiment NA60 at the CERN SPS has now achieved a decisive
breakthrough in this field.

\section{Apparatus and data analysis}
\label{sec:1}
The apparatus is based on the muon spectrometer previously used by
NA50, and a newly added telescope of radiation-tolerant silicon pixel
detectors, embedded inside a 2.5 T dipole magnet in the vertex
region~\cite{Gluca:2005,Keil:2005zq}. Matching of the muon tracks
before and after the hadron absorber, both in {\it angular and
momentum} space, improves the dimuon mass resolution in the region of
the light vector mesons from $\sim$80 to $\sim$20 MeV/c$^{2}$ and also
decreases the combinatorial background of muons from $\pi$ and K
decays. Moreover, the additional bend by the dipole field leads to a
strong increase of the detector acceptance for opposite-sign dimuons
of low mass and low transverse momentum. The rapidity coverage is
3.3$<$y$<$4.3 for the $\rho$, at low p$_{T}$ (compared to 3$<$y$<$4
for the J/$\psi$). Finally, the selective dimuon trigger and the fast
readout speed of the pixel telescope allow the experiment to run at
very high luminosities.

The results reported here were obtained from the analysis of data
taken in 2003 with a 158 AGeV indium beam, incident on a segmented
indium target of seven disks with a total of 18\% (In-In) interaction
length. At an average beam intensity of 5$\cdot$10$^{7}$ ions per 5~s
burst, about 3$\cdot$10$^{12}$ ions were delivered to the experiment,
and a total of 230 million dimuon triggers were recorded on tape. The
data reconstruction starts with the muon-spectrometer tracks. Next,
pattern recognition and tracking in the vertex telescope are done; the
interaction vertex in the target is reconstructed with a resolution of
$\sim$200 $\mu$m for the z-coordinate and 10-20 $\mu$m in the
transverse plane. Only events with one vertex are kept; interaction
pileup and reinteractions of secondaries and fragments are thus
rejected. Finally, each muon-spectrometer track is extrapolated to the
vertex region and matched to the tracks from the vertex telescope.

The combinatorial background of uncorrelated muon pairs mainly
originating from $\pi$ and K decays is determined using a {\it
mixed-event technique}~\cite{Ruben:2005qm}. Two single mu\-ons from
different like-sign dimuon triggers are combined into muon pairs in
such a way as to accurately account for details of the acceptance and
trigger conditions. The quality of the mixed-event technique can be
judged by comparing the like-sign distributions generated from mixed
events with the measured like-sign distributions. It is remarkable
that the two agree to within $\sim$1\% over a dynamic range of 4
orders of magnitude in the steeply falling mass
spectrum~\cite{Ruben:2005qm}. After subtraction of the combinatorial
background, the remaining opposite-sign pairs still contain ``signal''
fake matches, i.e. associations of muons to non-muon tracks in the
pixel telescope. This contribution is only 7\% of the combinatorial
background level. It has been determined in the present analysis by an
overlay Monte Carlo method. We have verified that an event-mixing
technique gives the same results, both in shape and in yield, within
better than~5\%. More details on the experimental apparatus and data
analysis will be given in a forthcoming extended paper; for now
see~\cite{Ruben:2005qm,Andre:2006}.

\section{Results}
\label{sec:2}
A significant part of the results presented in this paper has recently
been published~\cite{Arnaldi:2006}. That part will therefore be less
extensively treated than other parts published here for the first
time.
 
\begin{figure}[t!]
\centering
\resizebox{0.4\textwidth}{!}{%
\includegraphics*[clip=,bb = 6 12 560 665]{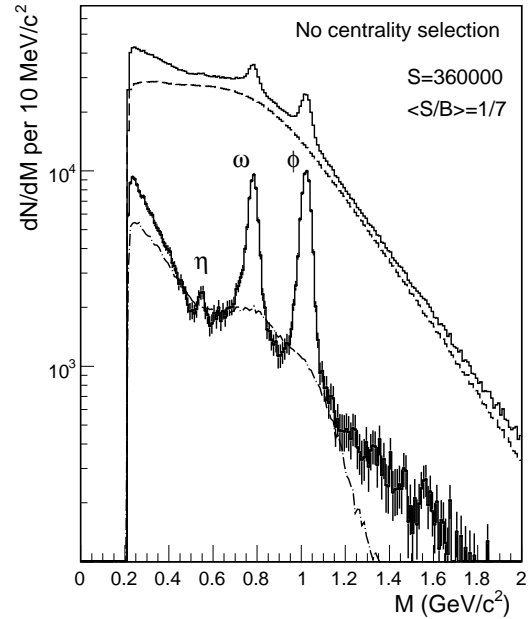}
}
\caption{Mass spectra of the opposite-sign dimuons (upper histogram),
combinatorial background (dashed), signal fake matches
(dashed-dotted), and resulting signal (histogram with error bars).}
\label{fig1}
\end{figure}

Fig.~\ref{fig1} shows the opposite-sign, background and signal dimuon
mass spectra, integrated over all collision centralities. After
subtracting the combinatorial background and the signal fake matches,
the resulting net spectrum contains about 360\,000 muon pairs in the
mass range 0-2 GeV/c$^2$ of Fig.~\ref{fig1}, roughly 50\% of the total
available statistics.  The average charged-particle multiplicity
density measured by the vertex tracker is $dN_{ch}/d\eta$ =120, the
average signal-to-background ratio is 1/7. For the first time in
nuclear collisions, the vector mesons $\omega$ and $\phi$ are
completely resolved in the dilepton channel; even the
$\eta$$\rightarrow$$\mu$$\mu$ decay is seen. The mass resolution at
the $\omega$ is 20 MeV/c$^{2}$. The subsequent analysis is done in
four classes of collision centrality defined through the
charged-particle multiplicity density: peripheral (4-30),
semiperipheral (30-110), semicentral (110-170) and central (170-240).
The signal-to-background ratios associated with the individual classes
are 2, 1/3, 1/8 and 1/11, respectively.

\begin{figure}
\includegraphics[width=0.45\textwidth]{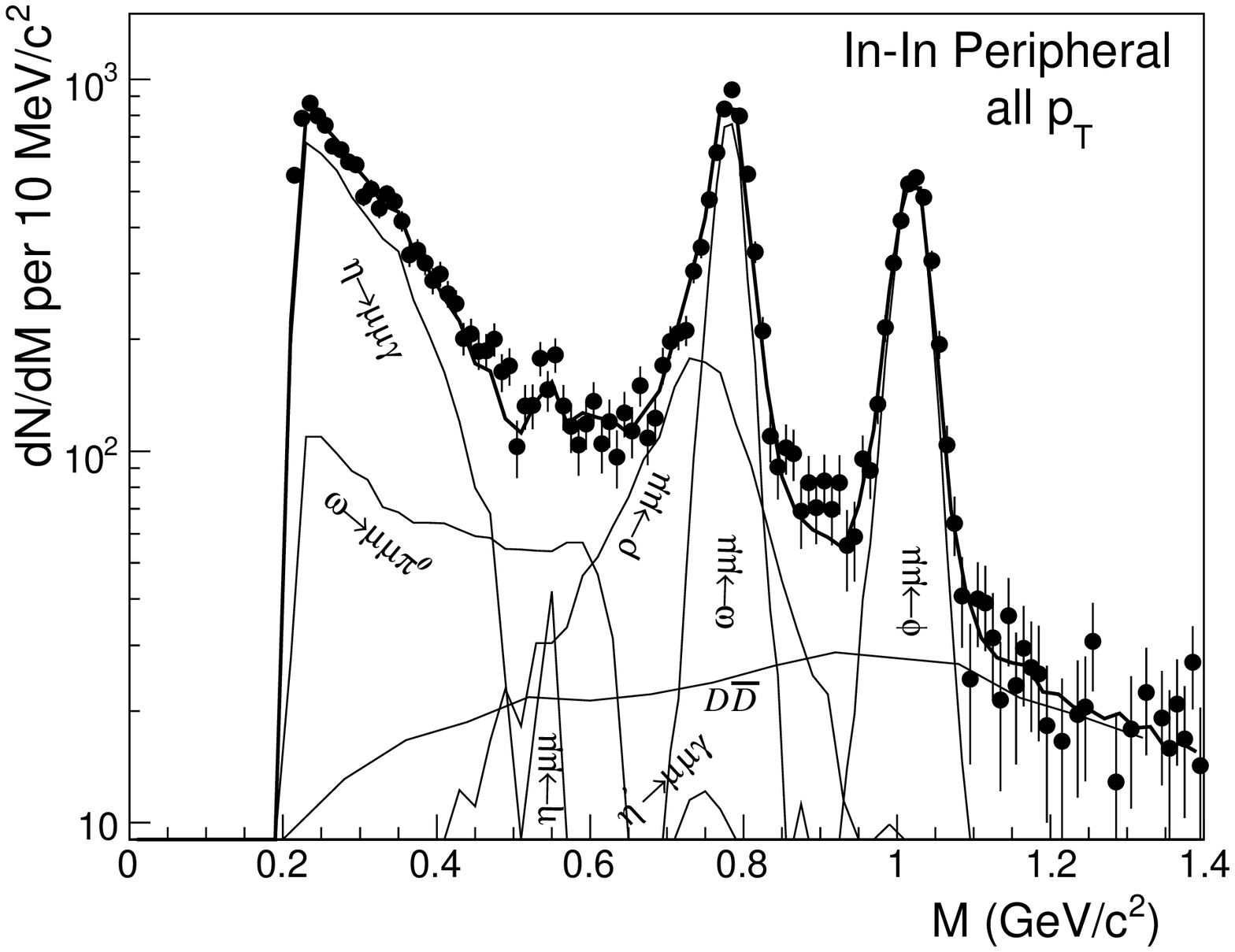}
\includegraphics[width=0.45\textwidth]{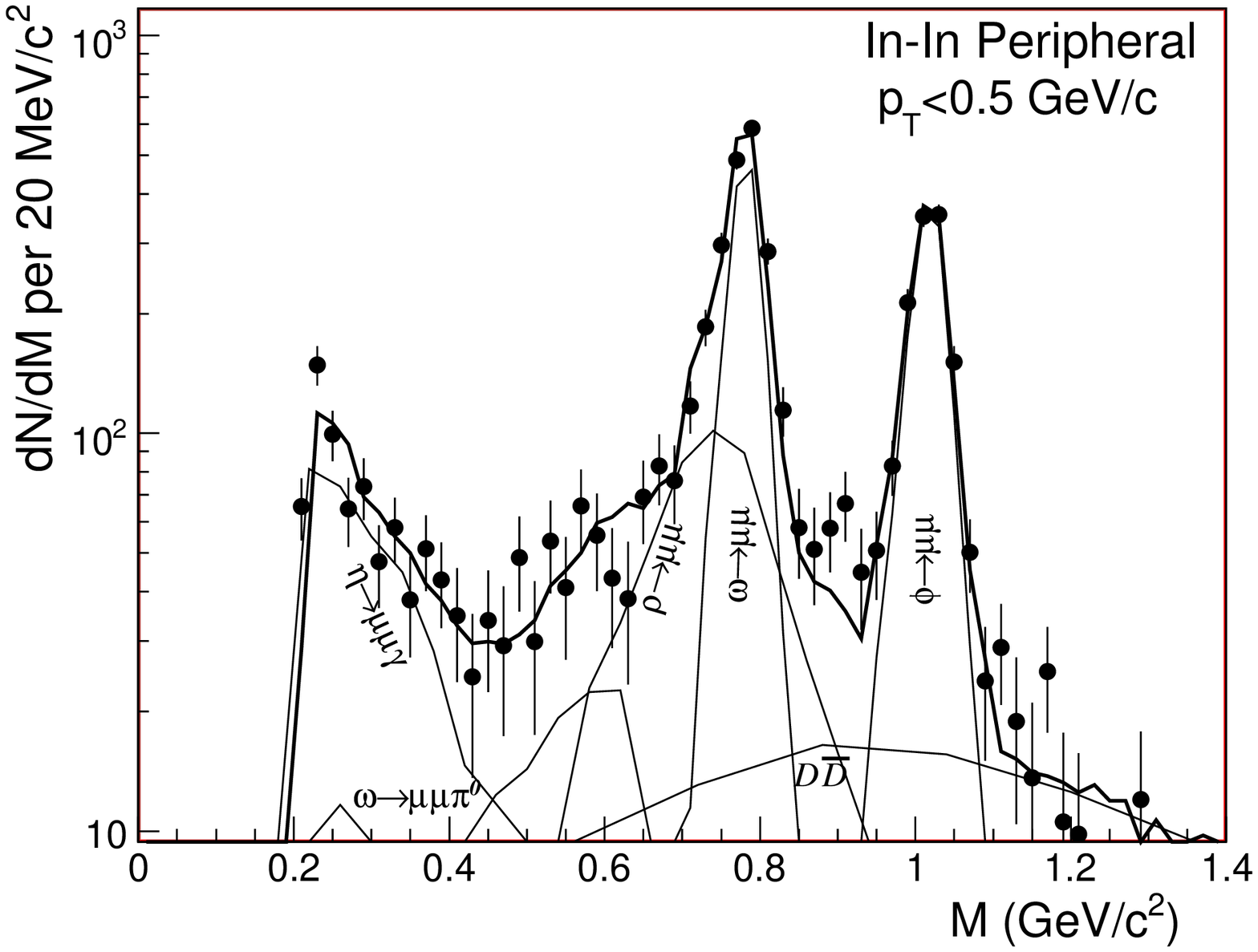}
  \caption{Fits of hadron decay cocktail to the peripheral data for
  all p$_{T}$ (upper) and p$_{T}$$<$0.5 GeV/c (lower), showing also
  the individual contributions.}
  \label{fig2}
\end{figure}

The peripheral data can essentially be described by the expected
electromagnetic decays of the neutral mesons. Muon pairs produced from
the 2-body decays of the $\eta$, $\rho$, $\omega$ and $\phi$
resonances and the Dalitz decays of the $\eta$, $\eta^{'}$ and
$\omega$
were simulated using the improved hadron decay generator
GENESIS~\cite{genesis:2003}, while GEANT was used for transport
through the detectors. Four free parameters (apart from the overall
normalization) were used in the fit of this ``hadron decay cocktail''
to the peripheral data: the cross section ratios $\eta/\omega$,
$\rho/\omega$ and $\phi/\omega$, and the level of $D$ meson pair
decays; the ratio $\eta^{'}/\eta$ was kept fixed at
0.12~\cite{Agakichiev:mv,genesis:2003}. The fits were done without
p$_{T}$ selection, but also independently in three windows of dimuon
transverse momentum: p$_{T}$$<$0.5, 0.5$<$p$_{T}$$<$1 and
p$_{T}$$>$1~GeV/c. Data and fits, including an illustration of the
individual sources, are shown in Fig.~\ref{fig2} for all p$_{T}$
(upper) and for the particular selection p$_{T}$$<$0.5 GeV/c
(lower). The fit quality is good throughout, even in the critical
acceptance-suppressed $\eta$-Dalitz region at low mass and low
p$_{T}$.

\begin{figure}
\includegraphics[width=0.45\textwidth]{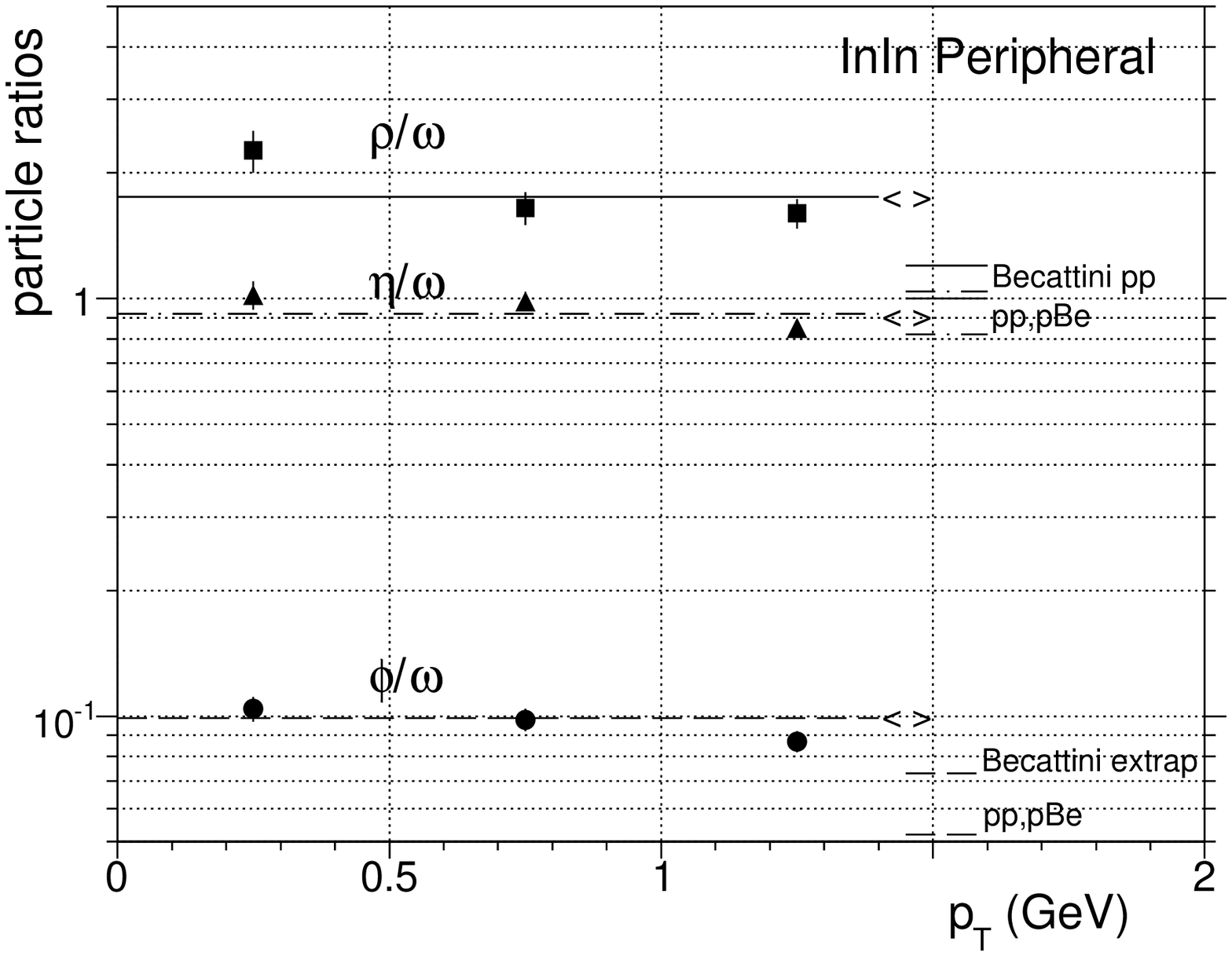}
\includegraphics[width=0.45\textwidth]{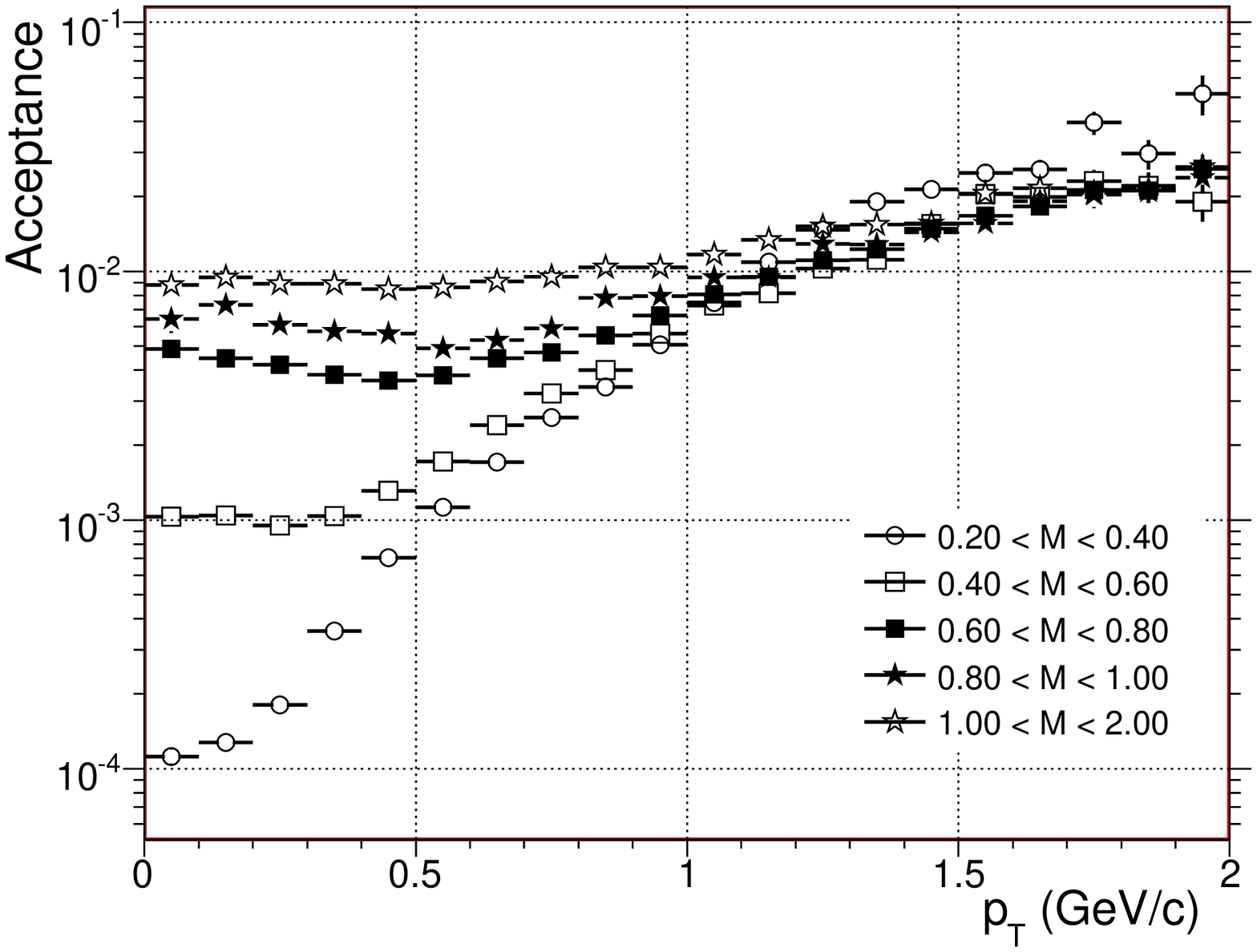}
  \caption{Upper: Particle cross-section ratios for all p$_{T}$ ($<>$)
  and for three p$_{T}$ windows, extrapolated to full phase
  space. Lower: NA60 acceptance relative to 4$\pi$ for different mass
  windows.}
  \label{fig3}
\end{figure}

The quantitative fit results in terms of the cross section ratios,
corrected for acceptance and extrapolated to full phase space (meaning
here full ranges in y and p$_{T}$) are displayed in Fig.~\ref{fig3}
(upper), where the horizontal lines (with the label $<>$) indicate the
fit values without p$_{T}$ selection, and the data points (with
statistical errors) the values obtained from the three different
p$_{T}$ windows. The systematic errors are of order 10\% in all cases,
dominated by those of the branching ratios. The $\eta/\omega$ ratio
agrees, within $<$10\%, with the literature average for p-p,
p-Be~\cite{Agakichiev:mv}. The $\phi/\omega$ ratio is higher than the
p-p, p-Be average, reflecting some $\phi$ enhancement already in
peripheral nuclear collisions. Both ratios are, within 10\%,
independent of the pair p$_{T}$. This implies that the GENESIS input
assumptions (used in the extrapolation to full p$_{T}$) as well as the
acceptance corrections vs. p$_{T}$ are correct on the level of
10\%. As shown in Fig.~\ref{fig3} (lower), the acceptance variations
with p$_{T}$ are minor for the $\omega$ and the $\phi$, but very
strong,
over two orders of magnitude, for the $\eta$ Dalitz mode (M$<$0.4
GeV/c$^{2}$). The accuracy level of 10\% reached in understanding the
acceptance is therefore truly remarkable.

The particle ratio $\rho/\omega$ behaves in a different way relative
to the other two. It decreases with p$_{T}$, but remains significantly
higher than the p-p, p-Be average~\cite{Agakichiev:mv}
throughout. This suggests that some $\pi\pi$ annihilation, enhancing
the yield of the $\rho$, contributes already in peripheral collisions
(see below).

\begin{figure}[t!]
\centering
\includegraphics*[width=7cm, height=7.8cm, clip=, bb= 0 12 560 665]{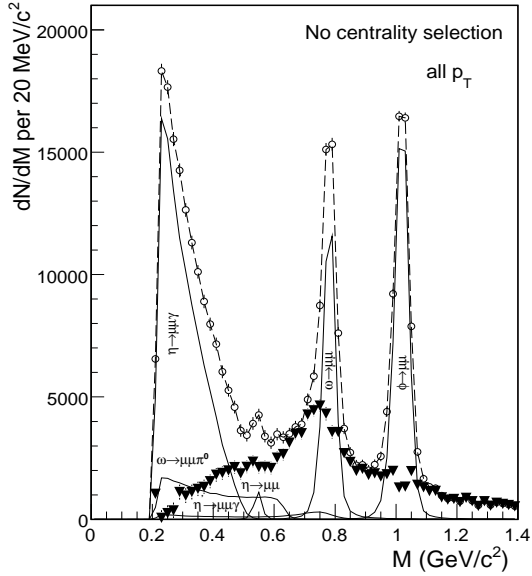}
\caption{Isolation of an excess above the hadron decay cocktail (see text). Total
data (open circles), individual cocktail sources (solid), difference
data (thick triangles), sum of cocktail sources and difference data
(dashed).}
\label{fig4}
\end{figure}

\begin{figure}[b!]
\vspace*{-0.5cm}
\centering
\resizebox{0.45\textwidth}{!}{%
\includegraphics*[width=4.7cm, height=3.7cm]{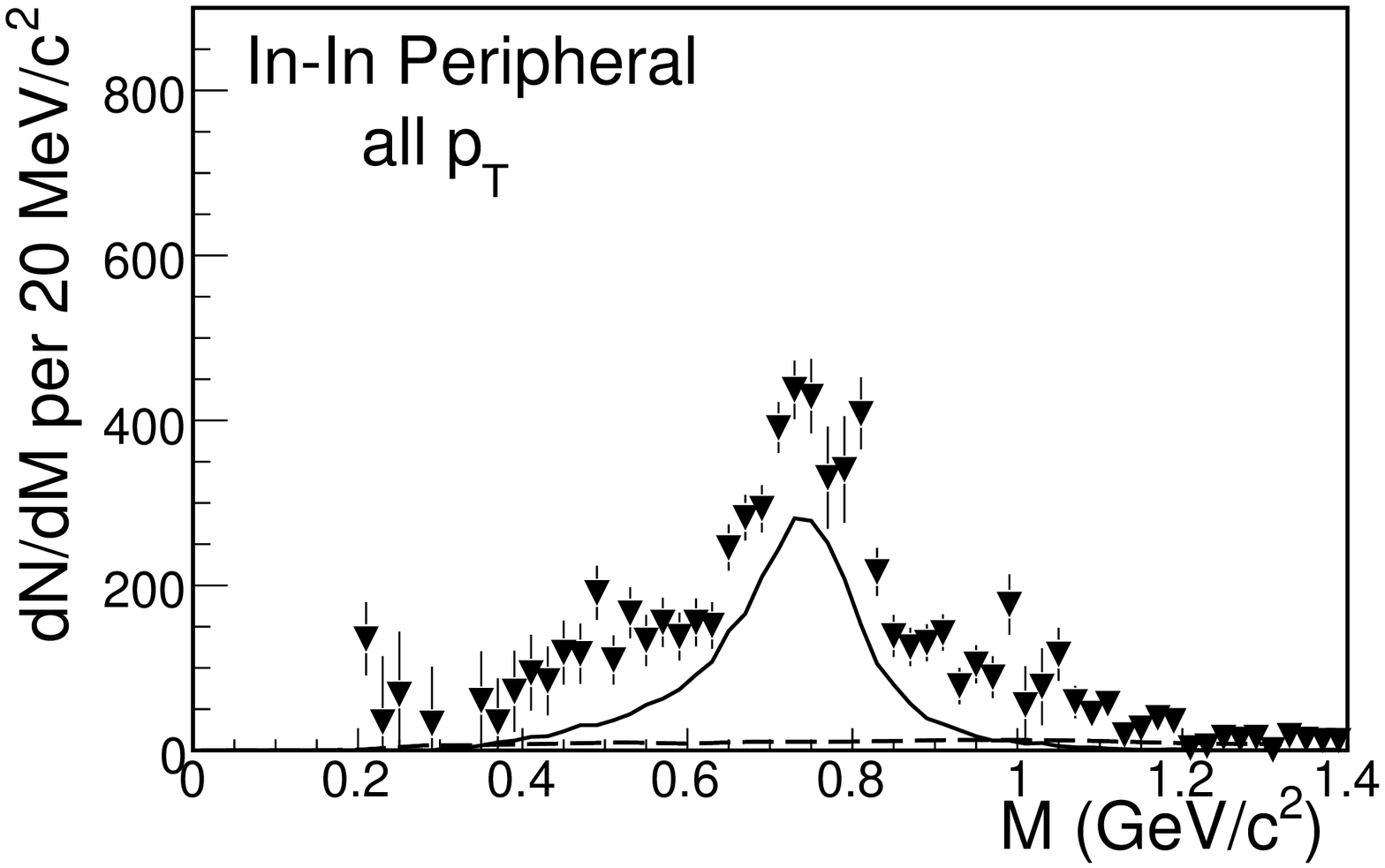}
\includegraphics*[width=4.7cm, height=3.7cm]{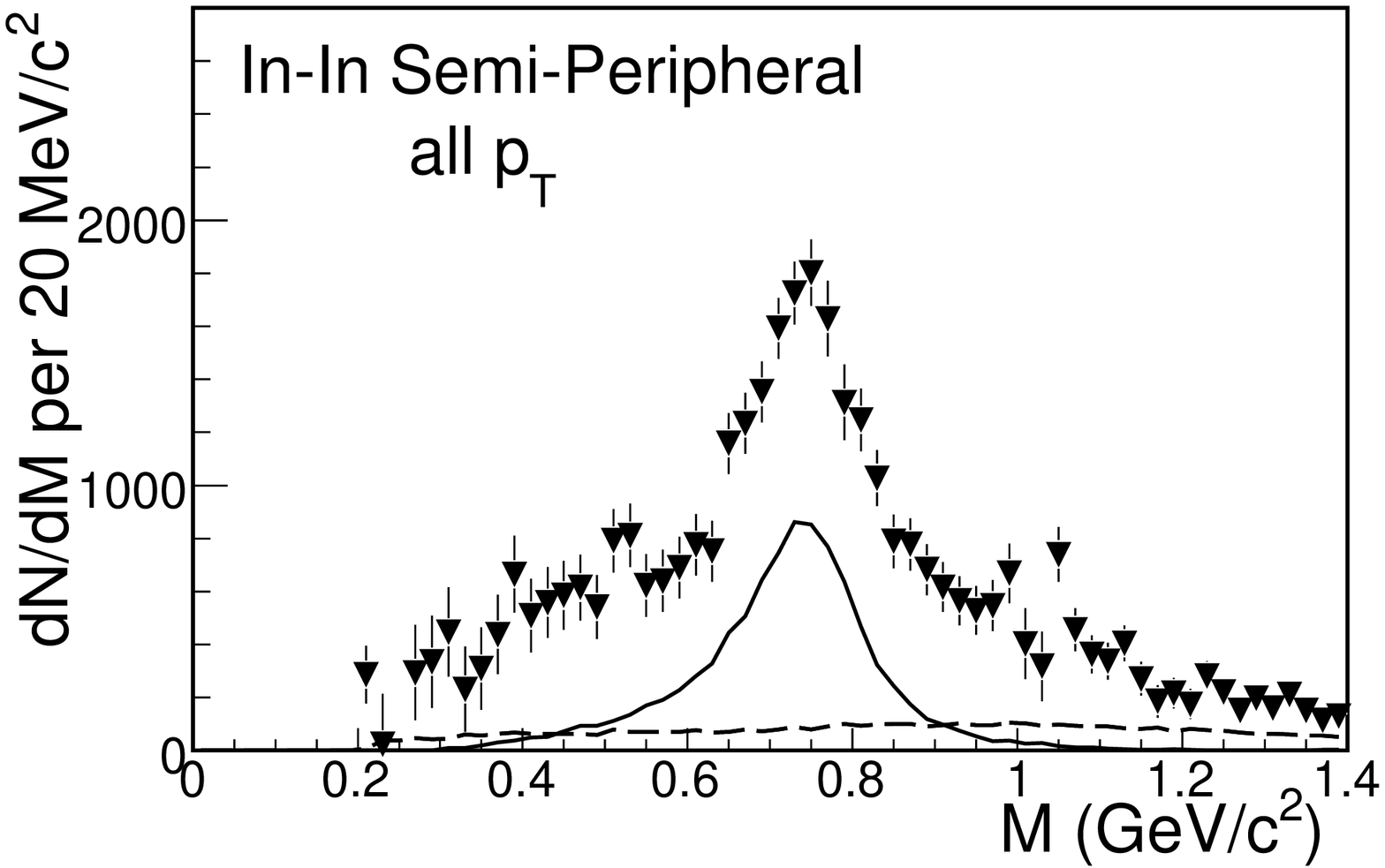}
}
\resizebox{0.45\textwidth}{!}{%
\includegraphics*[width=4.7cm, height=3.7cm]{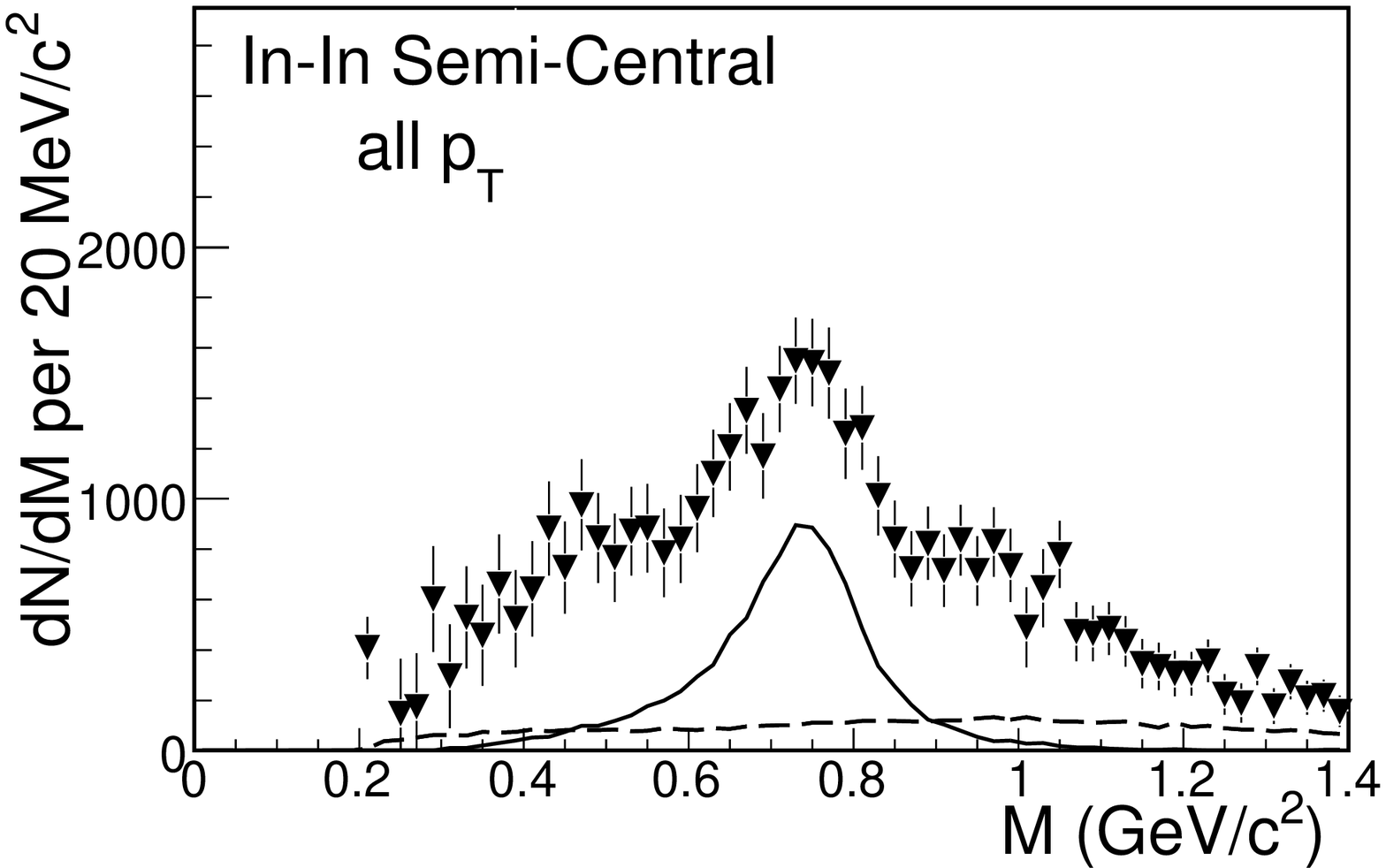}
\includegraphics*[width=4.7cm, height=3.7cm]{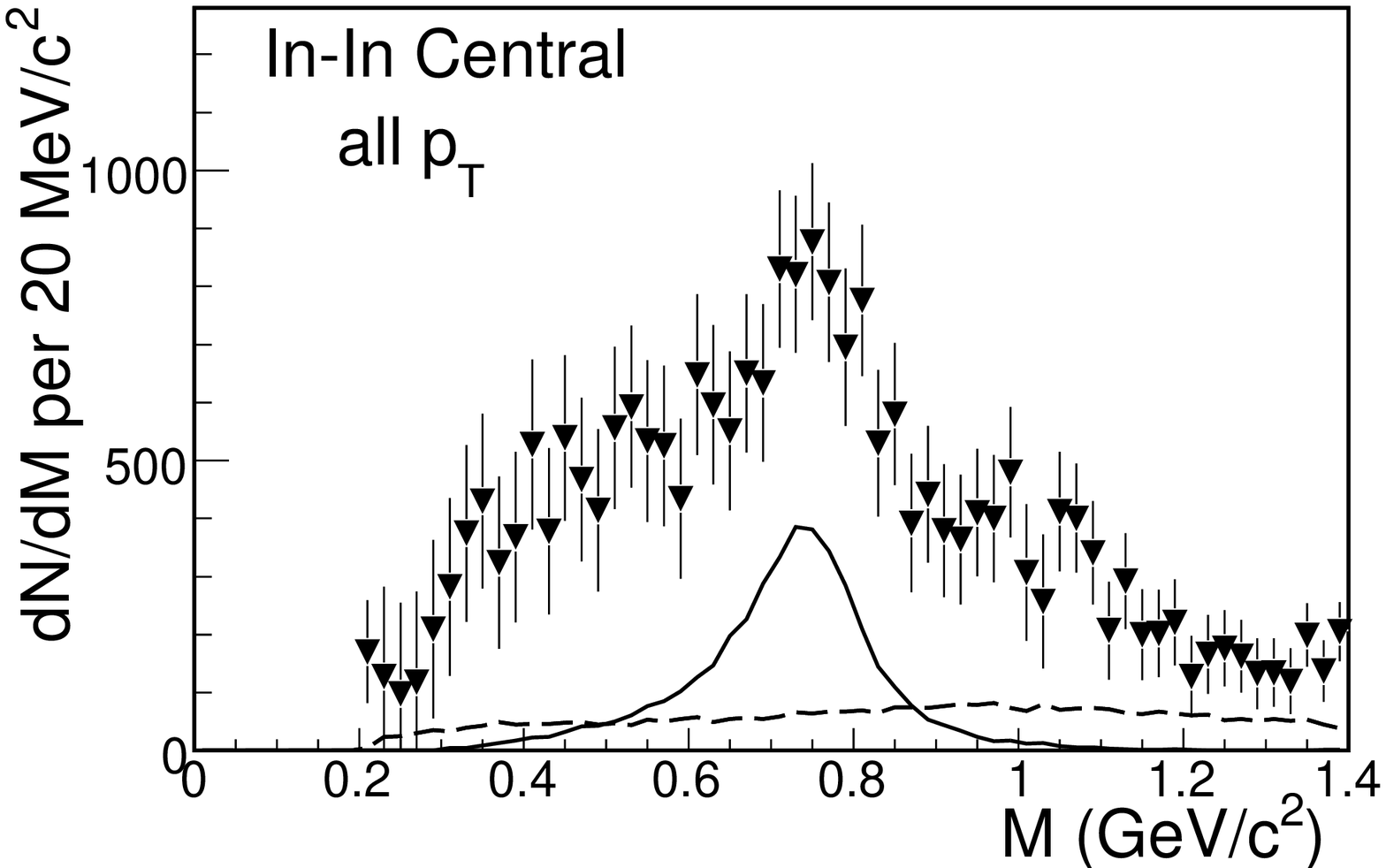}
}
\caption{Excess mass spectra of dimuons without p$_{T}$ selection. The
cocktail $\rho$ (solid) and the level of uncorrelated charm decays
(dashed) are shown for comparison. For errors see text.}
\label{fig5}
\end{figure}

\begin{figure}[b!]
\vspace*{-0.5cm}
\centering
\resizebox{0.45\textwidth}{!}{%
\includegraphics*[width=4.7cm, height=3.7cm]{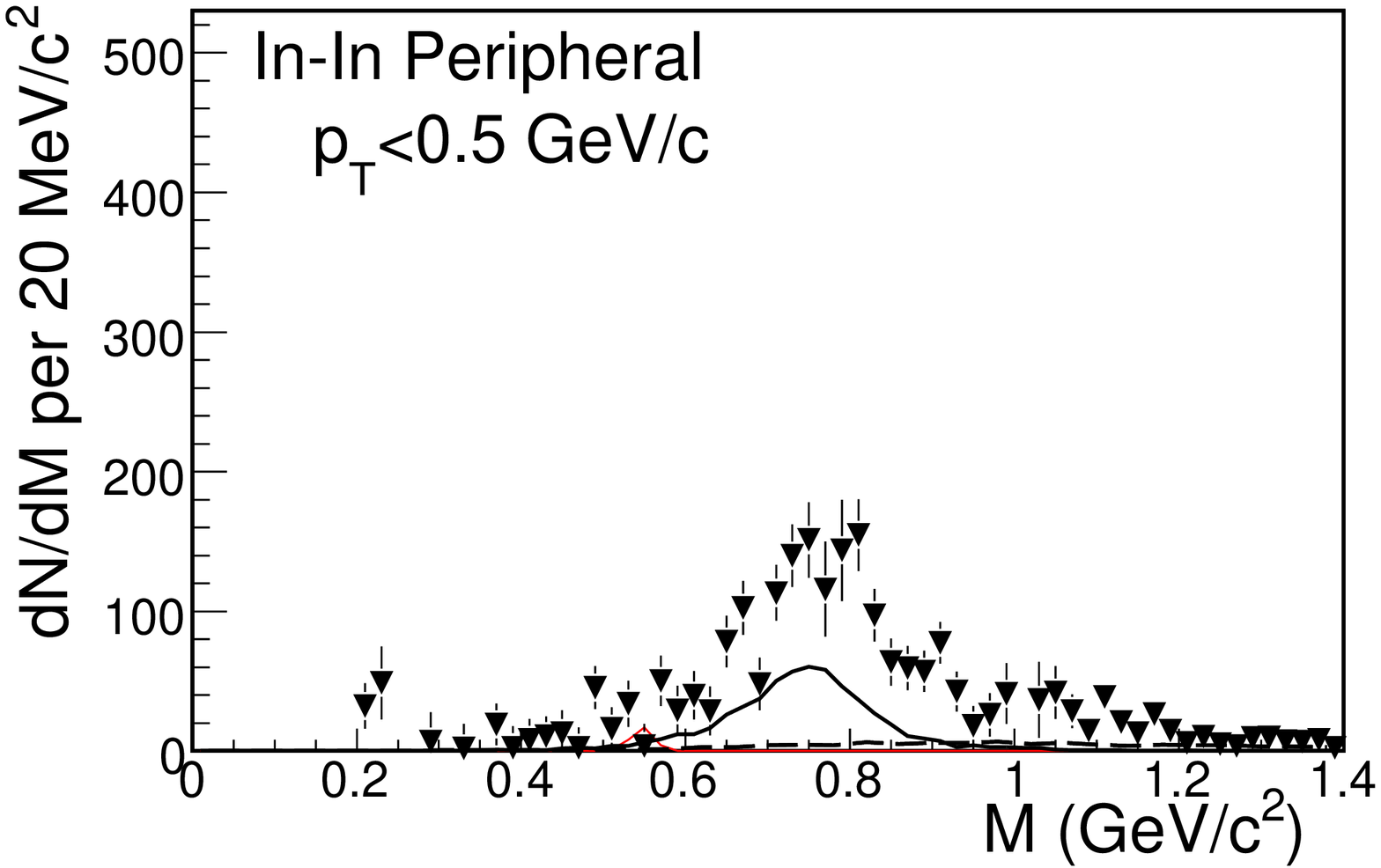}
\includegraphics*[width=4.7cm, height=3.7cm]{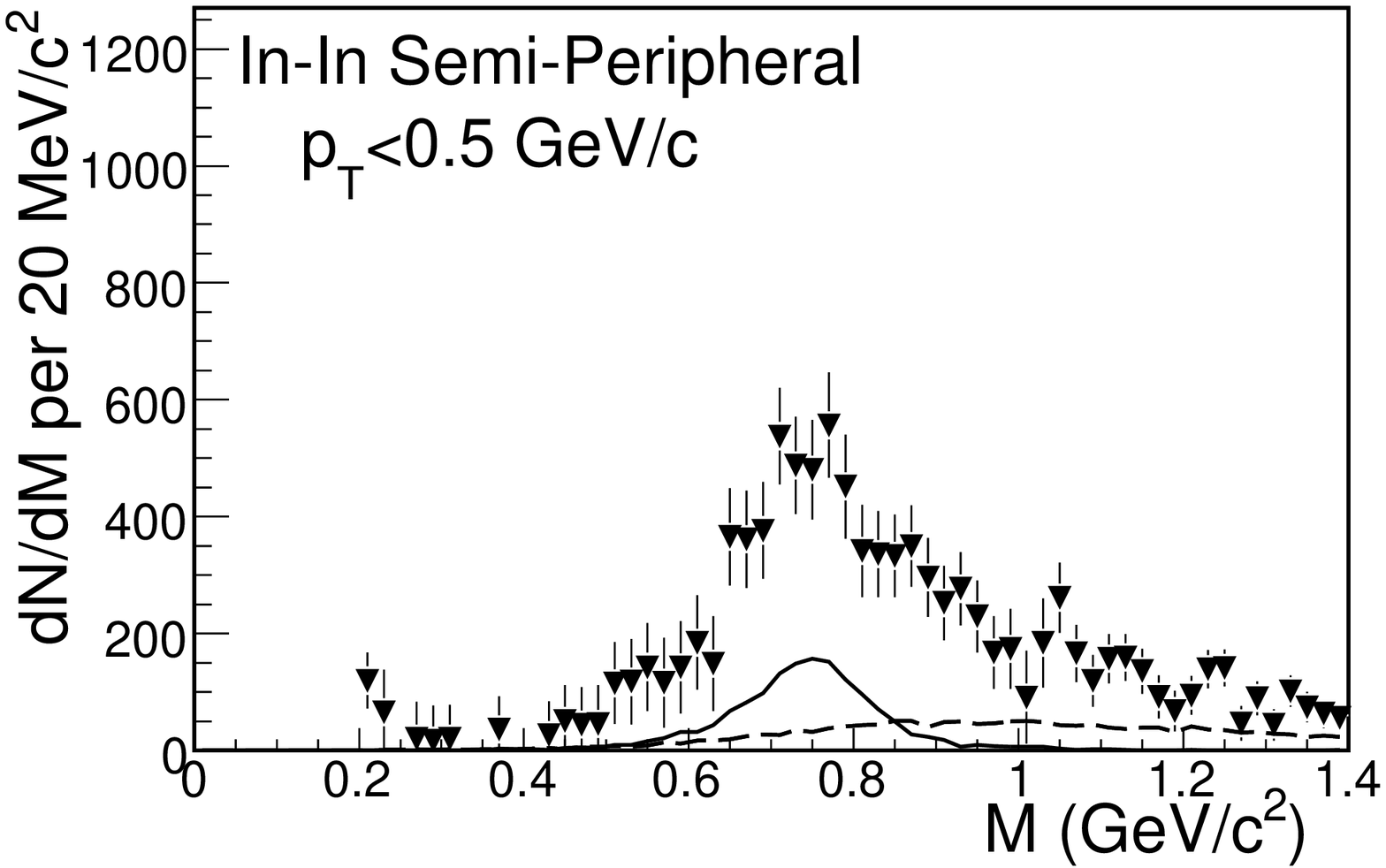}
}
\resizebox{0.45\textwidth}{!}{%
\includegraphics*[width=4.7cm, height=3.7cm]{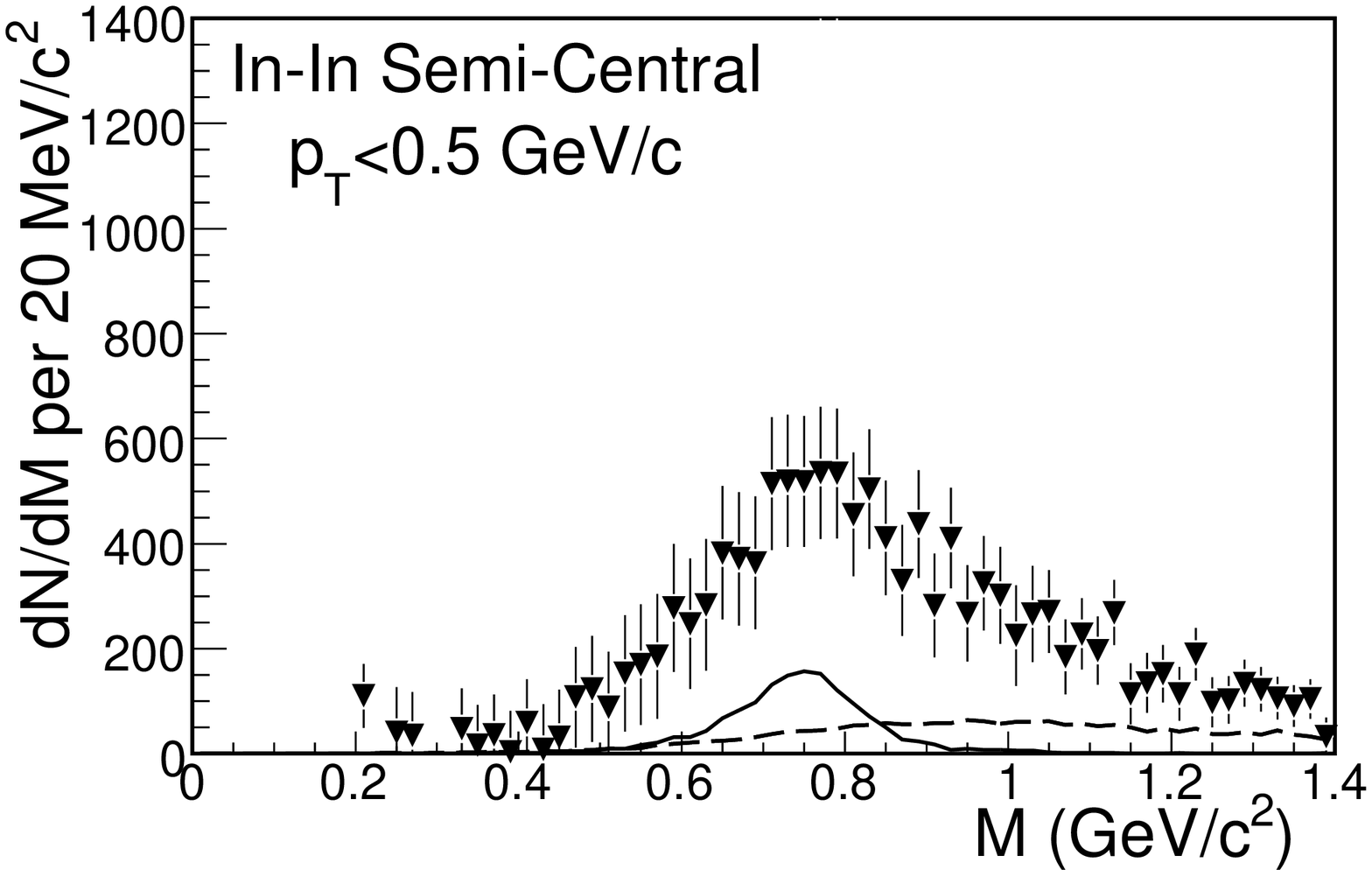}
\includegraphics*[width=4.7cm, height=3.7cm]{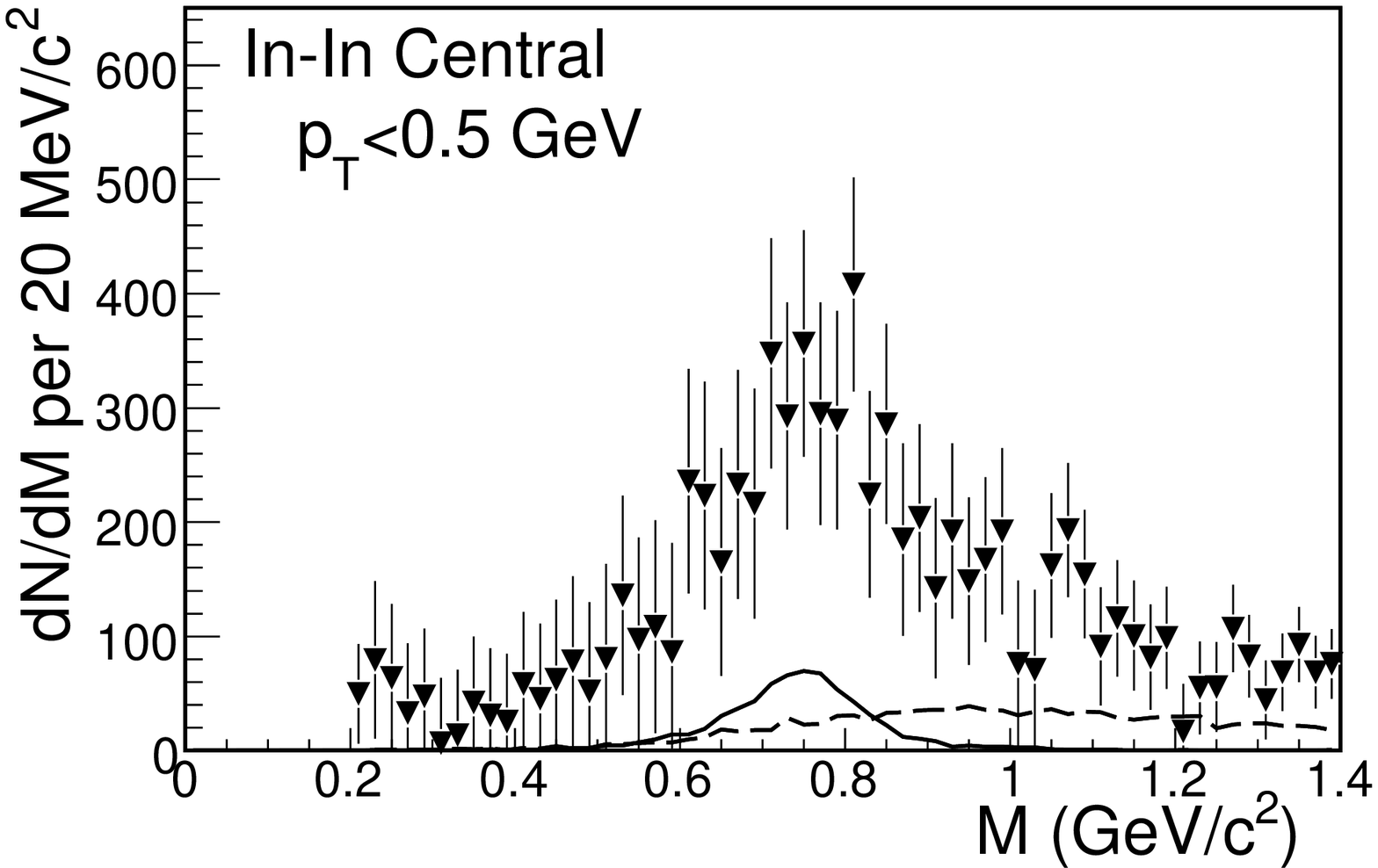}
}
\caption{Excess mass spectra of dimuons for p$_{T}$$<$0.5 GeV/c. The
cocktail $\rho$ (solid) and the level of uncorrelated charm decays
(dashed) are shown for comparison. For errors see text.}
\label{fig6}
\end{figure}

In the more central bins, a fit procedure is ruled out, due to the
existence of a strong excess with {\it a priori unknown}
characteristics. We have therefore used a novel procedure as shown in
Fig.~\ref{fig4}, made possible by the high data quality. The excess is
{\it isolated} by subtracting the cocktail, without the $\rho$, from
the data. The cocktail is fixed, separately for the major sources and
in each centrality bin, by a ``conservative'' approach.  The yields of
the narrow vector mesons $\omega$ and $\phi$ are fixed so as to get,
after subtraction, a {\it smooth} underlying continuum. For the
$\eta$, an upper limit is defined by ``saturating'' the measured data
in the region close to 0.2 GeV/c$^{2}$; this implies the excess to
vanish at very low mass, by construction.  The $\eta$ resonance and
$\omega$ Dalitz decays are now bound as well; $\eta^{'}/\eta$ is fixed
as before. The {\it cocktail $\rho$} (shown in
Figs.~\ref{fig5},~\ref{fig6},~\ref{fig8}~and~\ref{fig9} for
illustration purposes) is bound by the ratio $\rho/\omega$=1.2.  The
accuracy in the determination of the $\omega$ and $\phi$ yields by
this subtraction procedure is on the level of very few \%, due to the
remarkable {\it local} sensitivity, and not much worse for the $\eta$.

The excess mass spectra for all 4 multiplicity bins, resulting from
subtraction of the ``conservative'' hadron decay cocktail from the
measured data, are shown in Figs.~\ref{fig5} and \ref{fig6} for all
p$_{T}$ and the particular selection p$_{T}$$<$0.5 GeV/c,
respectively. The cocktail $\rho$ and the level of charm decays, found
in the three upper centrality bins to be about 1/3 of the measured
yield in the mass interval 1.2$<$M$<$1.4
GeV/c$^{2}$~\cite{Ruben:2005qm}, are shown for comparison. The
qualitative features of the spectra are striking: a peaked structure
is seen in all cases, broadening strongly with centrality, but
remaining essentially centered around the position of the nominal
$\rho$ pole.  At the same time, the total yield increases relative to
the cocktail $\rho$, their ratio (for M$<$0.9 GeV/c$^{2}$) reaching
values of 4 for all p$_{T}$, even close to 8 for p$_{T}$$<$0.5 GeV/c,
in the most central bin. Such values are consistent with the
CERES~\cite{Agakichiev:1997au} results, if the latter are also
referred
to the cocktail $\rho$ and rescaled to In-In. The errors shown are purely statistical. The
systematic errors are dominantly connected to the uncertainties in the
level of the combinatorial background, less so to the fake
matches. For the data without p$_{T}$ selection, they are estimated to
be about 3\%, 12\%, 25\% and 25\% for the 4 centralities in the broad
continuum region, while the $\rho$-like structure above the continuum
is much more robust. Uncertainties associated with the subtraction of
the hadron decay cocktail reach locally up to 15\%, dominated by those
of the $\omega$-Dalitz form factor. For the data with p$_{T}$$<$0.5
GeV/c, the systematic errors
are still under investigation.

\begin{figure}[t!]
\centering
\includegraphics[width=0.45\textwidth]{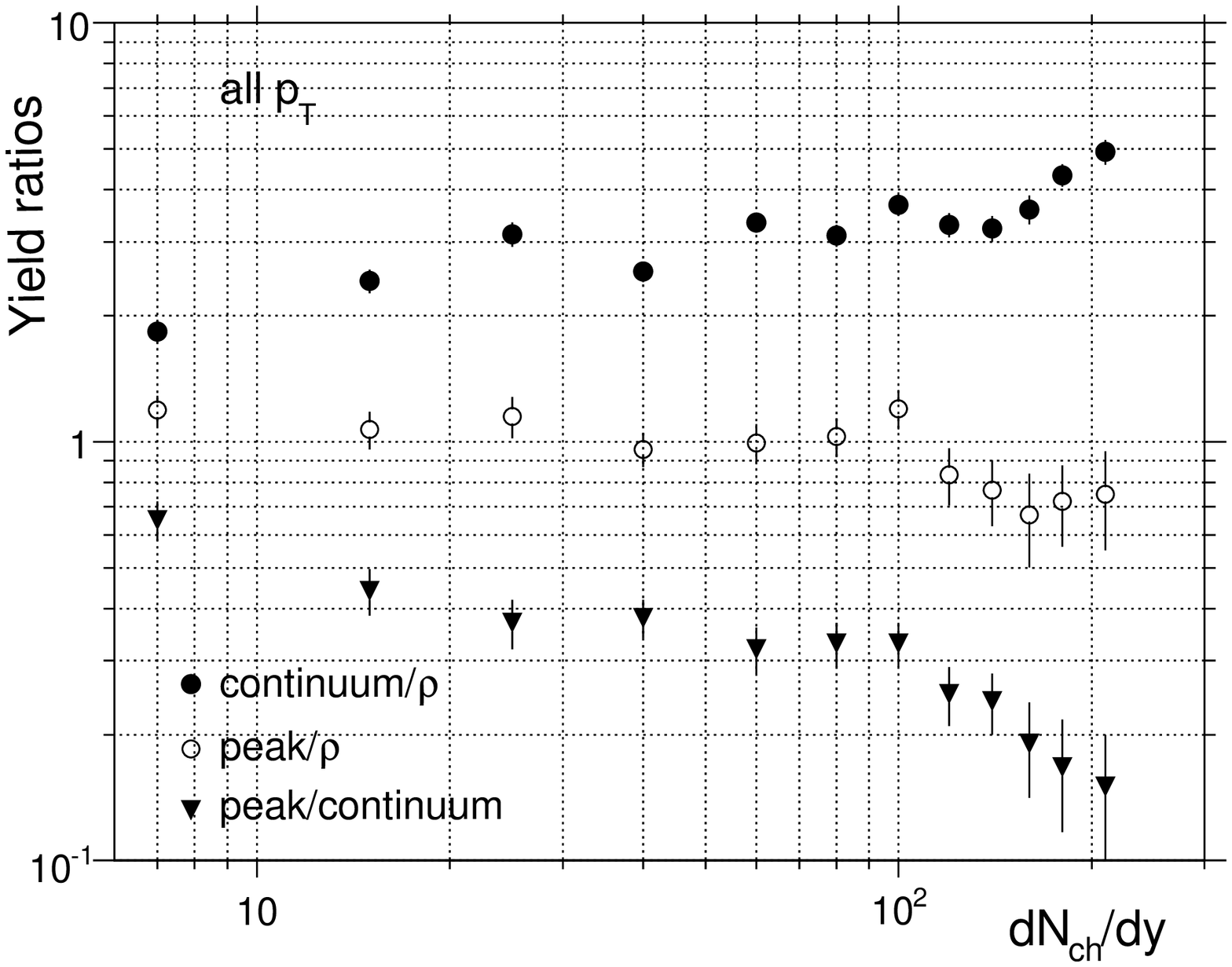}
\includegraphics[width=0.45\textwidth]{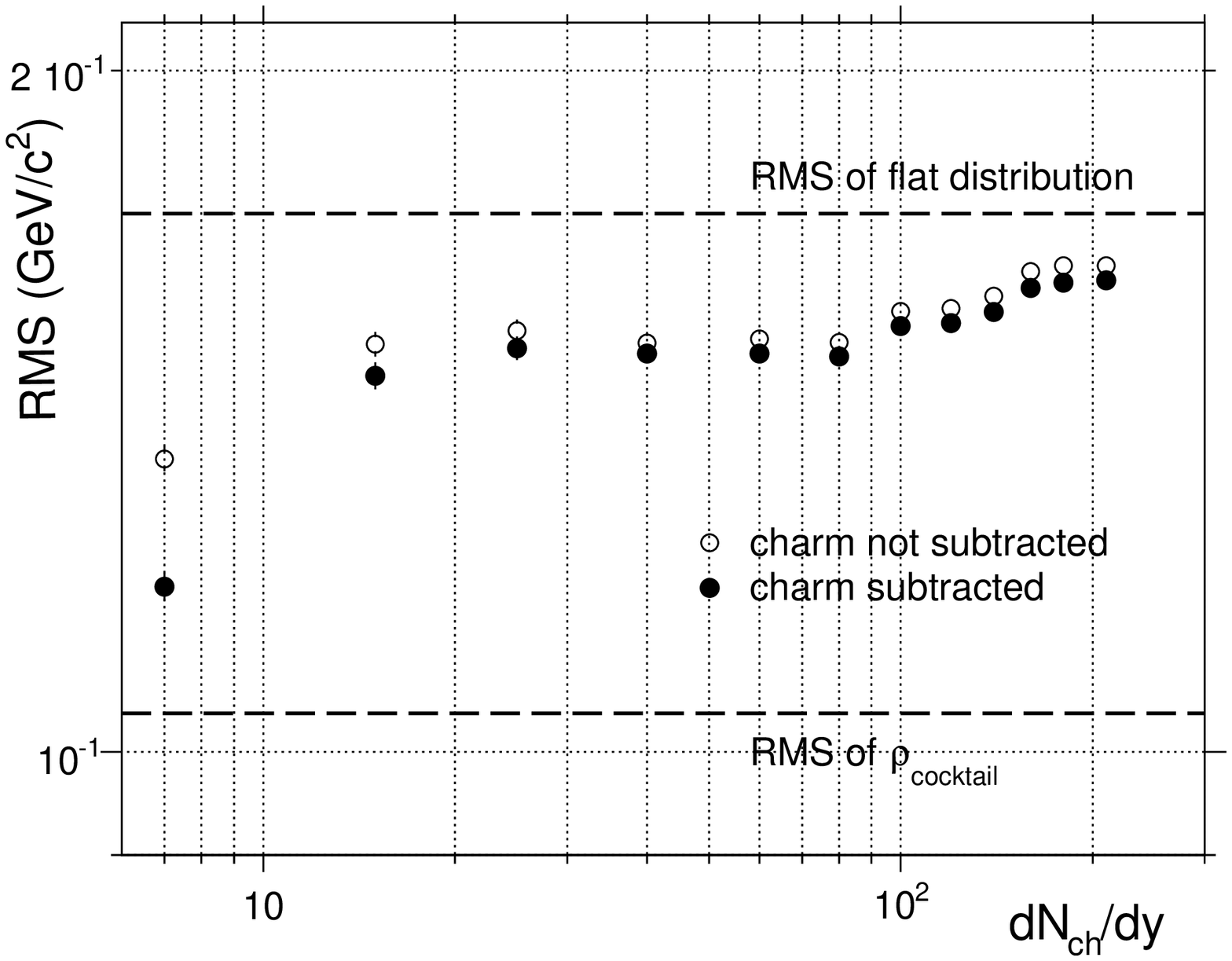}
\caption{Upper: Yield ratios continuum/$\rho$, peak/$\rho$, and
peak/continuum; see text for definition and errors. Lower: RMS of the
excess mass spectra in the window 0.44$<$M$<$1.04 GeV/c$^{2}$.}
\label{fig7}
\end{figure}

For the data without p$_{T}$ selection, a quantitative analysis of the
shape of the excess mass spectra vs. centrality has been performed,
using a finer subdivision into 12 centrality bins. Referring to
Fig.~\ref{fig5}, the data were subdivided into three mass windows with
equal widths: 0.44$<$M$<$0.64 (L=Lower), 0.64$<$M$<$0.84 (C=Center),
and 0.84$<$M$<$1.04 GeV/c$^{2}$ (U=Upper). From the yields in these
windows, a peak yield R = C - 1/2(L+U) and a continuum yield 3/2(L+U)
can be defined. Fig.~\ref{fig7} (upper) shows the ratios of
peak/$\rho$ (R/$\rho$), continuum/$\rho$ (3/2(L+U)/$\rho$), and
peak/continuum (RR), where $\rho$ stands for the cocktail $\rho$. The
errors shown are purely statistical.  The relative systematic errors
between neighboring points are small compared to the statistical
errors and can therefore be ignored, while the common systematic
errors are related to those discussed in the previous paragraph and
are of no relevance here. The ratio peak/$\rho$ is seen to decrease
from the most peripheral to the most central bin by nearly a factor of
2, ruling out the naive view that the shape can simply be explained by
the cocktail $\rho$ residing on a broad continuum, independent of
centrality. The ratio continuum/$\rho$ shows a fast initial rise,
followed by a more flat and then another more rapid rise beyond
dN$_{ch}$/dy = 100; this behavior is statistically significant. The
sum of the two ratios is the total enhancement factor relative to the
cocktail $\rho$; it reaches about 5.5 in the most central bin. The
ratio of the two ratios, peak/continuum, amplifies the two separate
tendencies: a fast decay, a nearly constant part, and a decline by a
further factor of 2 beyond dN$_{ch}$/dy = 100, though with large
errors due to R.

A completely independent shape analysis was done by just evaluating
the RMS = $\sqrt{\langle M^{2}\rangle-\langle M\rangle^{2}}$ of the
mass spectra in the single total mass interval 0.44$<$M$<$1.04
GeV/c$^{2}$. The results are shown in Fig.~\ref{fig7} (lower), both
for the full data and after subtraction of charm on the level
discussed for Figs.~\ref{fig5},~\ref{fig6} (except for the most
peripheral bin where the continuum yield above 1 GeV/c$^{2}$ was
assumed to be 100\% charm).
They are in perfect qualitative agreement with the more
``microscopic'' shape analysis discussed before, rising from values
close to the cocktail $\rho$ all the way up to nearly a flat-continuum
value. The extra rise beyond dN$_{ch}$/dy = 100 is highly significant
here, due to the very small statistical and systematic errors.

\section{Comparison to theoretical models}
\label{sec:3}

\begin{figure}[b!]
\vspace*{-0.5cm}
\centering
\includegraphics[width=0.45\textwidth]{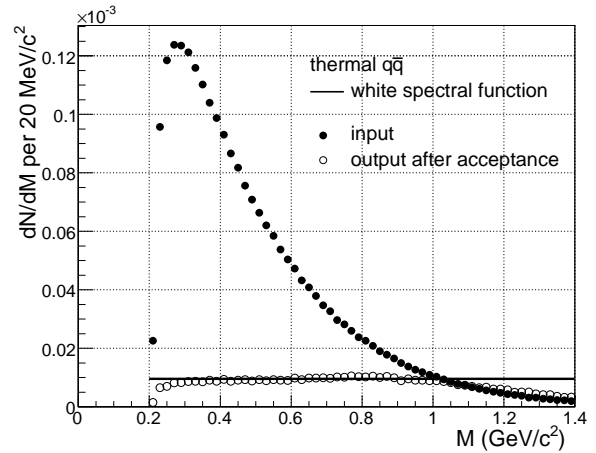}
\caption{Propagation of thermal $q\bar{q}$ radiation, based on a
uniform spectral function, through the NA60 acceptance without any
p$_{T}$ selection. The resulting spectrum is, {\it by coincidence},
also uniform.}
\label{fig8}
\end{figure}

The qualitative features of the excess mass spectra shown in
Figs.~\ref{fig5}~and~\ref{fig6} are consistent with an interpretation
as direct thermal radiation from the fireball, dominated by $\pi\pi$
annihilation. A quantitative comparison of the data to the respective
theoretical models can either be done at the {\it input} of the
experiment, requiring acceptance correction of the data, or at the
{\it output}, requiring propagation of the theoretical results through
the experimental acceptance.  All the data contained in this paper
have so far not been corrected for acceptance, and therefore only the
second alternative is available at present (the first one being under
preparation). To help intuition, Fig.~\ref{fig8} illustrates the
effects of acceptance propagation for the particularly transparent
case of $q\bar{q}$ annihilation, associated with a uniform spectral
function~\cite{rapp:nn23}. {\it By coincidence}, without p$_{T}$
selection, the resulting mass spectrum at the {\it output} is also
uniform within 10\% up to about 1.0 GeV/c$^{2}$, resembling the shape
of the spectral function at the {\it input}. In other words, the
always existing steep rise of the theoretical input at low masses
(Fig.~\ref{fig8}), due to the photon propagator and a Boltzmann-like
factor~\cite{Rapp:1995zy,Rapp:1999ej,Brown:kk,Brown:2001nh}, is just
about compensated by the falling acceptance in this region as long as
no p$_{T}$ cut is applied. Variations of the input p$_{T}$ spectrum
within reasonable physics limits affect the flatness of the output by
at most 20\%. Strong cuts in p$_{T}$ like $<$0.5 or $>$1 GeV/c,
however, completely invalidate the argument.

On the basis of this discussion, the excess mass spectra shown in
Fig.~\ref{fig5} can approximately be interpreted as spectral functions
of the $\rho$, averaged over momenta and the complete space-time
evolution of the fireball. The broad continuum-like part of the
spectra may then reflect the early history close to the QCD boundary
with a nearly divergent width, while the narrow peak on top may just
be due to the late part close to thermal freeze-out, approaching the
nominal width. The p$_{T}$-cut data shown in Fig.~\ref{fig6}, on the
other hand, do not allow such interpretation, due to the extreme
acceptance cut on the low-mass side of the $\rho$; compare also
Fig.~\ref{fig3} (lower).

Among the many different theoretical predictions for the properties of
the intermediate $\rho$ mentioned in the introduction, only few have
been brought to a level suitable for a quantitative comparison to the
data. Before the first release of the present data in 2005, the
in-medium broadening scenario of~\cite{Rapp:1995zy,Rapp:1999ej} and
the moving mass scenario related to~\cite{Brown:kk,Brown:2001nh} were
evaluated for In-In at dN$_{ch}$/d$\eta$=140, 
\begin{figure*}[]
\resizebox{1.\textwidth}{!}{%
\includegraphics*{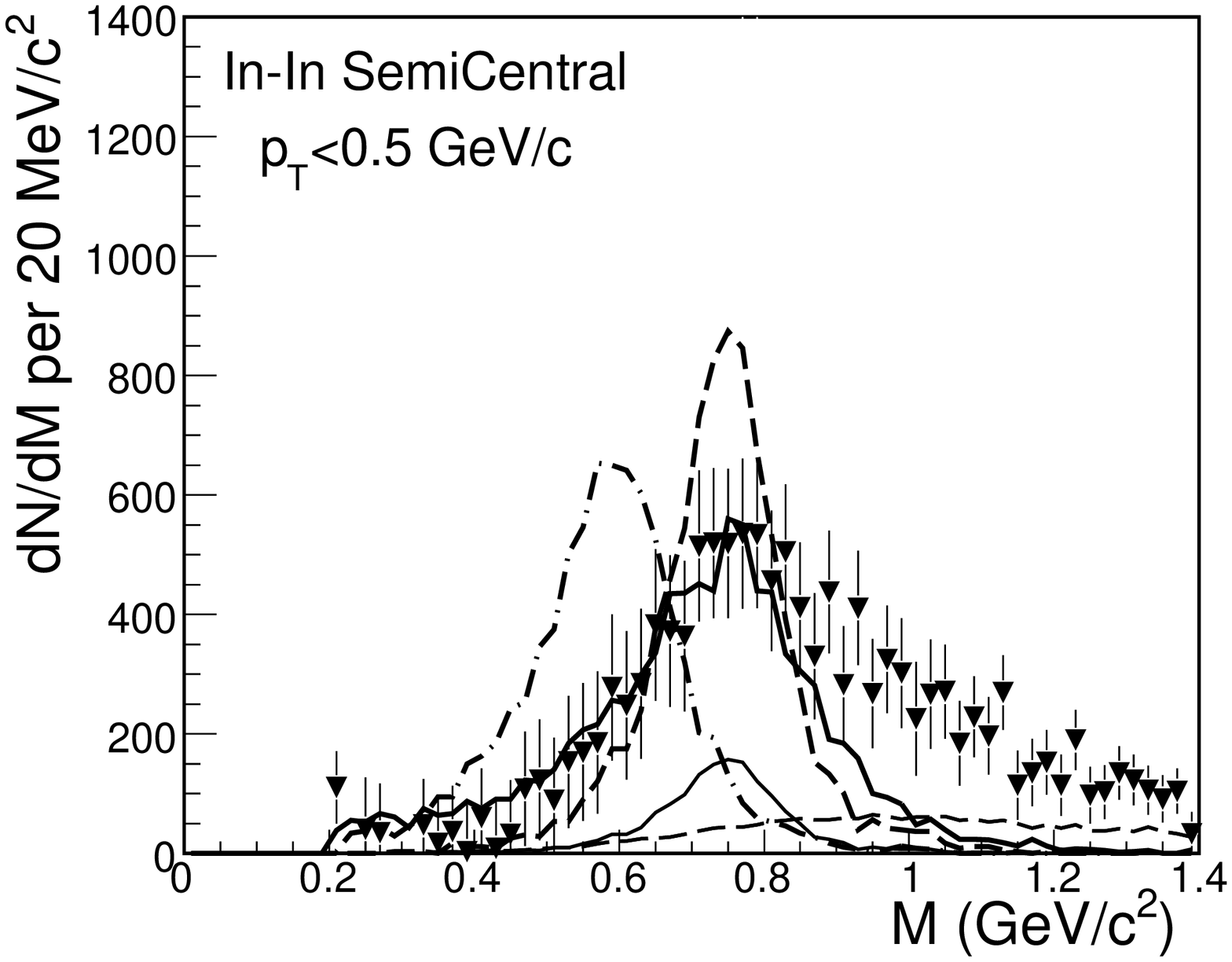}
\includegraphics*{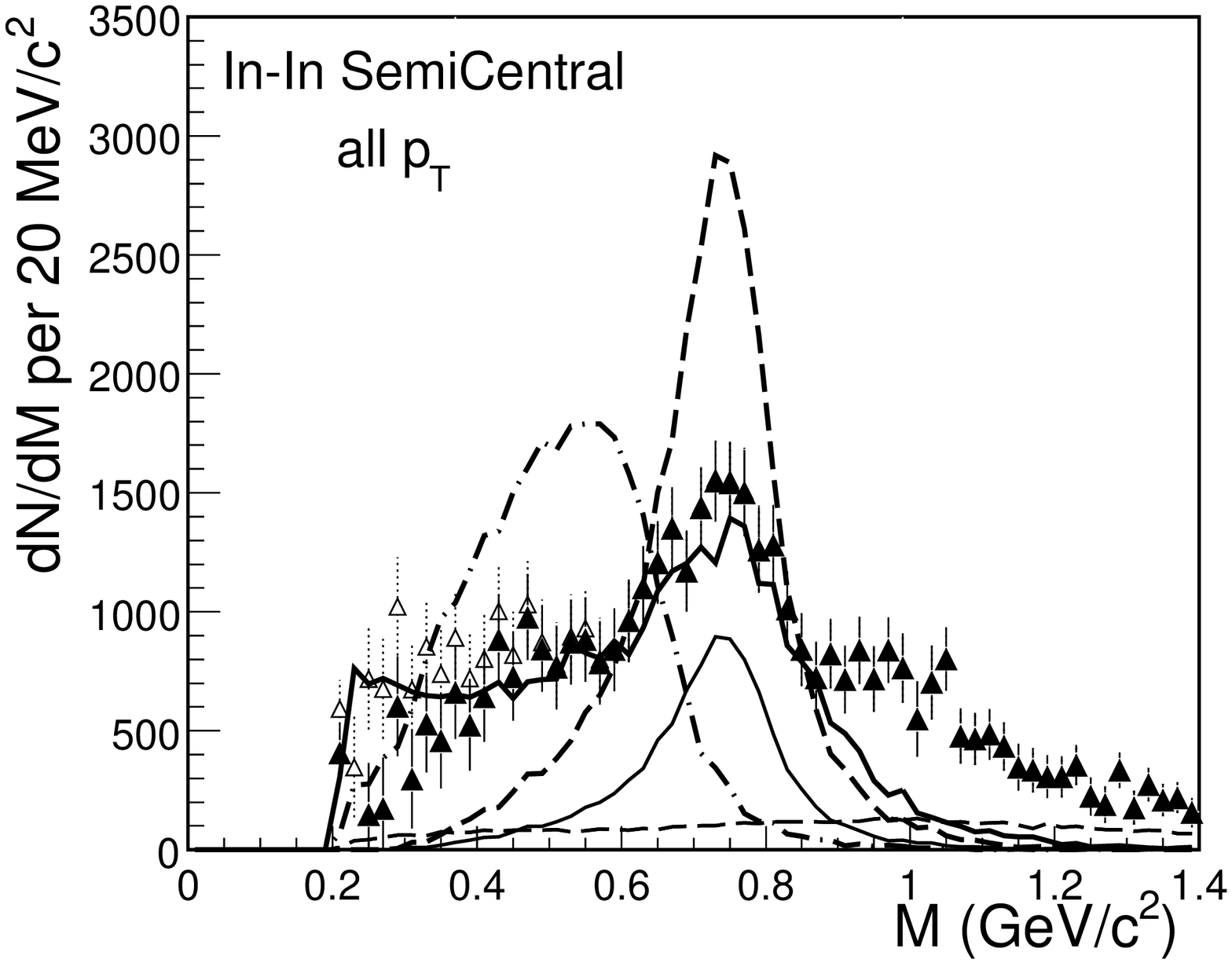}
\includegraphics*{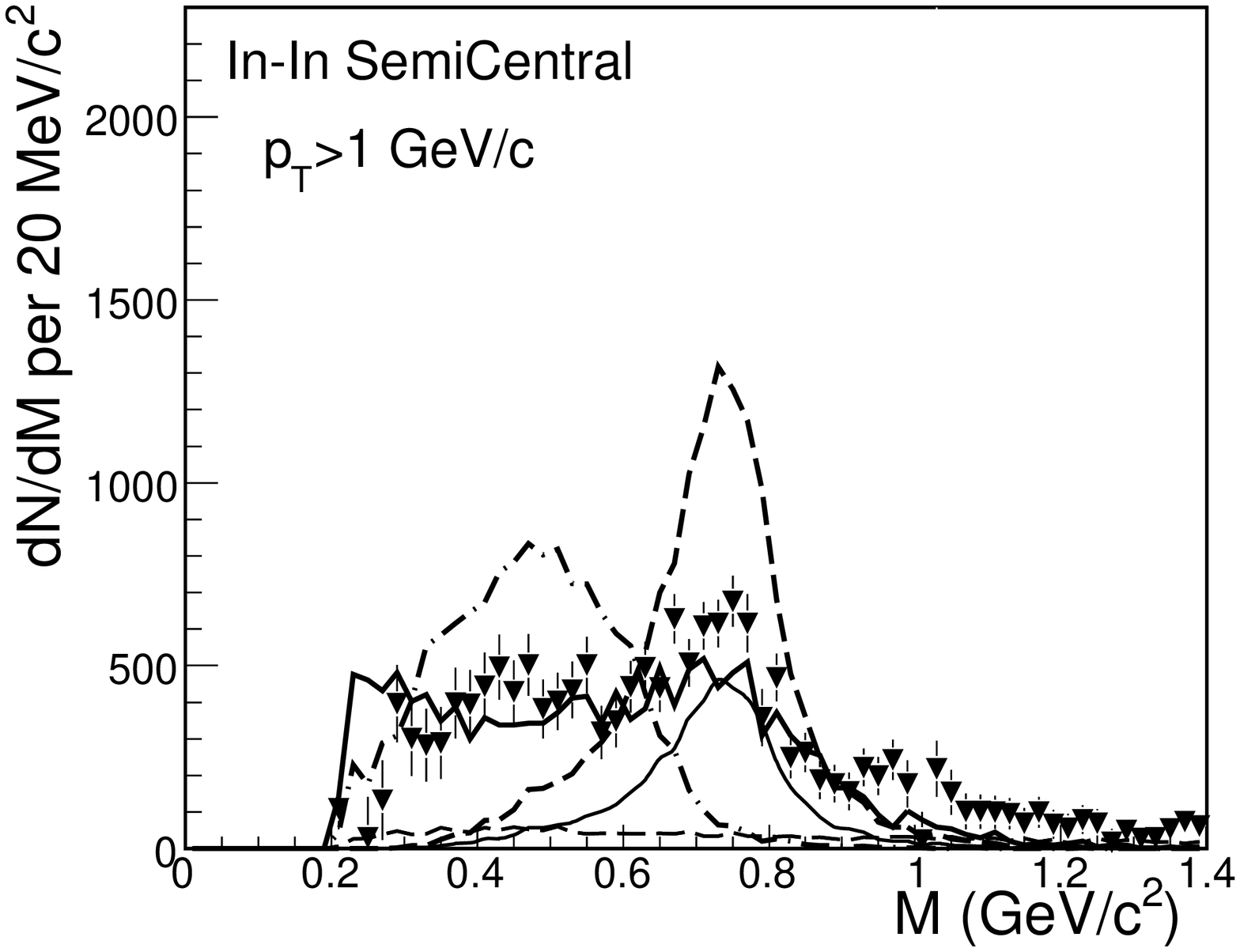}
}
\caption{Comparison of excess mass spectra for p$_{T}$$<$0.5, all
p$_{T}$ and $p_{T}$$>$1 GeV/c to model predictions, made for In-In at
$dN_{ch}/d\eta$=140 (semicentral bin). Cocktail $\rho$ (thin solid),
unmodified $\rho$ (dashed), in-medium broadening
$\rho$~\cite{Rapp:1995zy,Rapp:1999ej} (thick solid), in-medium moving
$\rho$ related to~\cite{Brown:kk,Brown:2001nh} (dashed-dotted),
uncorrelated charm (thin dashed). The errors are purely statistical;
for systematic errors see text. The open data points show the
difference spectrum resulting from a decrease of the $\eta$ yield by
10\% (which should also be viewed as a systematic error).}
\label{fig9}
\end{figure*}
using exactly the same fireball evolution model extrapolated from
Pb-Pb, and taking explicit account of temperature as well as of baryon
density~\cite{rapp:nn23}. In Fig.~\ref{fig9}, these predictions (as
well as the unmodified $\rho$) are confronted with the data for the
semicentral bin. The integrals of the theoretical spectra are
independently normalized to the data in the mass interval M$<$0.9
GeV/c$^{2}$ in order to concentrate on the spectral shapes,
independent of the uncertainties of the fireball evolution. Note again
that data and predictions can only be interpreted as space-time
averaged spectral functions for the part without p$_{T}$ selection
(Fig.~\ref{fig9}, center), but not for the other two. Irrespective of
the choice of the p$_{T}$ window, however, some general conclusions
can be drawn. The unmodified $\rho$ is clearly ruled out. The moving
mass scenario related to Brown/Rho scaling, which fit the CERES
data~\cite{Rapp:1999ej,Agakichiev:1997au}, is also ruled out, showing
the much improved discrimination power of the present data. Only the
broadening scenario appears to be in fair agreement with the data.

In the meantime, the release of the data in 2005 has triggered a
number of new theoretical developments. Contrary to initial
critics~\cite{brownrho:nn}, Brown/Rho scaling could not be saved by
varying fireball and other parameters within extremes, including
switching out the effects of temperature
altogether~\cite{rapphees:nn,skokov:nn}. The excess of the data at
M$>$0.9 GeV/c$^{2}$ may be related to the prompt dimuon excess found
by NA60 in the intermediate mass region~\cite{Ruben:2005qm}. It is not
accounted for by the model results shown in Fig.~\ref{fig9}, and it is
presently (nearly quantitatively) described either by hadronic
processes like 4$\pi$, 6$\pi$.. (including vector-axial-vector
mixing)~\cite{hees:nn}, or by partonic processes like $q\bar{q}$
annihilation~\cite{rr:nn}. This is a challenging theoretical ambiguity
to be solved in the future. A chiral virial approach has also been
able to nearly quantitatively describe the data~\cite{zahed:nn}.

\section{Conclusions}
\label{sec:5}

The data unambiguously show that the $\rho$ primarily broadens in
In-In collisions, but does not show any shift in mass. Consequently,
model comparisons favor broadening scenarios, but tend to rule out
moving-mass scenarios coupled directly to the chiral condensate. The
issue of vector-axialvector mixing, also sensitive to chiral
restoration, remains somewhat open at present. We expect that precise
p$_{T}$ dependences, presently under investigation, will give more
insight into the different sources operating in different mass
regions.


\begin{thebibliography}{}
\bibitem{Pisarski:mq} R.~D.~Pisarski, Phys.\ Lett.\ {\bf 110B}, 155
(1982)
\bibitem{Dominguez:1992dw}
  C.~A.~Dominguez, M.~Loewe and J.~C.~Rojas,
  Z.\ Phys.\ {\bf C59}, 63 (1993)
\bibitem{Pisarski:1995xu}
  R.~D.~Pisarski, Phys.\ Rev.\ D {\bf 52}, R3773 (1995)
\bibitem{Rapp:1995zy}
G.~Chanfray, R.~Rapp and J.~Wambach, Phys.~Rev. Lett.~{\bf 76}, 368 (1996);
R.~Rapp, G.~Chanfray and J.~Wambach, Nucl.\ Phys.\ {\bf A617}, 472 (1997)
\bibitem{Rapp:1999ej}
R.~Rapp and J.~Wambach, Adv.\ Nucl.\ Phys. {\bf 25}, 1 (2000)
\bibitem{Brown:kk}
G.~E.~Brown, M.~Rho, Phys.\ Rev.\ Lett. {\bf 66}, 2720 (1991);
G.~Q.~Li, C.~M.~Ko and G.~E.~Brown, Phys.\ Rev.\ Lett. {\bf 75}, 4007 (1995)
\bibitem{Brown:2001nh}
G.~E.~Brown and M.~Rho,
Phys.\ Rept. {\bf 363}, 85 (2002)
\bibitem{Hatsuda:1991ez}
T.~Hatsuda and S.~H.~Lee,
Phys.\ Rev.\ C {\bf 46}, 34 (1992)
\bibitem{Agakichiev:mv} G.~Agakichiev {\em et al.} (CERES Collaboration), 
Eur.\ Phys.\ J.\ {\bf C4}, 231 (1998)
\bibitem{Agakichiev:1995xb}
  G.~Agakichiev {\it et al.} (CERES Collaboration),
  Phys.\ Rev.\ Lett.\  {\bf 75}, 1272 (1995)
\bibitem{Agakichiev:1997au} G.~Agakichiev {\em et
al.} (CERES Collaboration), Phys.\ Lett.\ {\bf B422}, 405 (1998); B.~Lenkeit {\em et al.},
Nucl.\ Phys. {\bf A661}, 23c (1999); G.~Agakichiev {\it et al.}, Eur.\
Phys.\ J.\ {\bf C41}, 475 (2005)
\bibitem{Gluca:2005}
  G.~Usai {\it et al.} (NA60 Collaboration),
  Eur.\ Phys.\ J.\  {\bf C43}, 415 (2005)
\bibitem{Keil:2005zq}
  M.~Keil {\it et al.}, Nucl.\ Instrum.\ Meth.\ {\bf A539}, 137 (2005) 
  and {\bf A546}, 448 (2005)
\bibitem{Ruben:2005qm}
R.~Shahoyan {\em et al.} (NA60 Collaboration), 
Eur.\ Phys.\ J.\ C43 (2005) 209; Quark Matter, Budapest, 2005
\bibitem{Andre:2006}
A.~David, PhD Thesis, Instituto Superior T\'ecnico, Lisbon, 2006; CERN-THESIS-2006-007
\bibitem{Arnaldi:2006}
R.~Arnaldi {\it et al.}  (NA60 Collaboration), Phys.\ Rev.\ Lett.\  {\bf 96} (2006) 162302
\bibitem{genesis:2003} 
S.~Damjanovic, A.~De~Falco and H. W\"ohri (NA60 Collaboration), NA60 Internal Note 2005-1
\bibitem{rapp:nn23}
R.~Rapp, private communication (2003)
\bibitem{brownrho:nn}
G.~E.~Brown and M.~Rho, arXiv:nucl-th/0509001 and nucl-th/0509002
\bibitem{rapphees:nn}
H.~van Hees and R.~Rapp, arXiv:hep-ph/0604269
\bibitem{skokov:nn}
V.~V.~Skokov and V.~D.~Toneev, Phys.\ Rev.\ C {\bf 73} (2006) 021902
\bibitem{hees:nn}
H.~van Hees and R.~Rapp, arXiv:hep-ph/0603084
\bibitem{rr:nn}
T.~Renk and J.~Ruppert, arXiv:hep-ph/0605130.
\bibitem{zahed:nn}
 K.~Dusling,~D.~Teaney~and~I.~Zahed, arXiv:nucl-th/0604071
\end{thebibliography}
\end{document}